\documentclass[aps, prb, twocolumn, superscriptaddress, showpacs, 10pt]{revtex4-1}

\usepackage{graphicx}
\usepackage{dcolumn}
\usepackage{bm}
\usepackage{amsmath}
\usepackage{multirow}

\usepackage{color}
\usepackage[normalem]{ulem}

\begin{document}
\title{Heat flow in InAs/InP heterostructure nanowires}
\author{J.~Matthews}
\affiliation{Physics Department and Materials Science Institute, University of Oregon, Eugene, Oregon 97403-1274}
\author{E.~A.~Hoffmann}
\altaffiliation{Present address: Institute for Advanced Study, Technische Universit\"{a}t
M\"{u}nchen, Lichtenbergstrasse 2a, D-85748 Garching, Germany}
\affiliation{Physics Department and Materials Science Institute, University of Oregon, Eugene, Oregon 97403-1274}
\author{C.~Weber}
\affiliation{Mathematical Physics and Nanometer Structure Consortium (nmC@LU), Lund University, Box 118, S-221 00, Lund, Sweden}
\author{A.~Wacker}
\affiliation{Mathematical Physics and Nanometer Structure Consortium (nmC@LU), Lund University, Box 118, S-221 00, Lund, Sweden}
\author{H.~Linke}
\affiliation{Physics Department and Materials Science Institute, University of Oregon, Eugene, Oregon 97403-1274}
\affiliation{Solid State Physics and Nanometer Structure Consortium (nmC@LU), Lund University, Box 118, S-221 00, Lund, Sweden}

\newcommand{\expandTable}[0]{\raisebox{0pt}[9pt][2pt]}

\begin{abstract}
The transfer of heat between electrons and phonons plays a key role for thermal management in future nanowire-based devices, but only a few experimental measurements of electron-phonon (e-ph) coupling in nanowires are available. Here, we combine experimental temperature measurements on an InAs/InP heterostructure nanowire system with finite element modeling (FEM) to extract information on heat flow mediated by e-ph coupling. We find that the electron and phonon temperatures in our system are highly coupled even at temperatures as low as 2 K. Additionally, we find evidence that the usual power-law temperature dependence of electron-phonon coupling may not correctly describe the coupling in nanowires and show that this result is consistent with previous research on similar one-dimensional electron systems. We also compare the strength of the observed e-ph coupling to a theoretical analysis of e-ph interaction in InAs nanowires, which predicts a significantly weaker coupling strength than observed experimentally.
\end{abstract}


\maketitle
\section{Introduction}
There exists great potential for the use of nanowires (NW) for future nanoelectronic applications, such as light-emitting diodes (LEDs)\cite{Bao2006, Bavencove2010}, photovoltaic (PV) cells\cite{Kelzenberg2008, Garnett2010}, wrap-gate transistors\cite{Bryllert2006, Egard2010}, and low-dimensional thermoelectrics\cite{Zhou2005, Dresselhaus2007, Boukai2008}. In many of these devices heat flow plays a significant role in device performance, either because heat flow is a parasitic effect and is therefore undesired, as is the case in thermoelectrics, or because high heat flow is critically required for thermal management, as in most other applications, such as LEDs, PV cells and transistors. In many cases, heat is produced in the form of Joule heat in the electronic system, but is then distributed between the electrons and phonons through electron-phonon coupling. Since phonons generally have the higher thermal conductivity in semiconductors, electron-phonon coupling has a central role in heat flow through semiconductor nanowires. 

Few experimental papers discuss e-ph interaction in semiconductor nanowire systems, where electrons and phonons are confined to one-dimensional transport at low temperatures.\cite{Kent1999, Sugaya2002, Prasad2004, Prasad2004_2} On the theory side, e-ph interaction strength was previously analyzed using three-dimensional (3D) phonon systems, which typically leads to the prediction of $T^3$ or $T^5$ power laws depending on the type of coupling considered.\cite{Bockelmann1990, Kubakaddi2007} However, in experimental systems with 1D electrons and 3D phonons, experimentally observed  power laws do not agree with these predictions.\cite{Sugaya2002, Prasad2004, Prasad2004_2} In vertically grown NWs then, where both electrons and phonons can be confined to 1D, one would expect the same, if not greater, deviation from the usual power laws. In fact, many previous theoretical analyses of phonon confinement in nanoscale structures have predicted increased e-ph interaction energy exchange and scattering rates compared to bulk systems \cite{Yu1994, Svizhenko1998, Jin2006, Ramayya2008, Vartanian2008, Luisier2009, Zhang2010}, deviations from the $T^3$ or $T^5$ power laws\cite{Vartanian2008}, and e-ph interaction regimes where powers laws are not obeyed\cite{DasSarma1993}. Together, these previous results highlight the need for additional investigation of e-ph coupling in nanowires, both experimentally and from a fundamental point of view.

\begin{figure}[t]
\begin{center}
\includegraphics{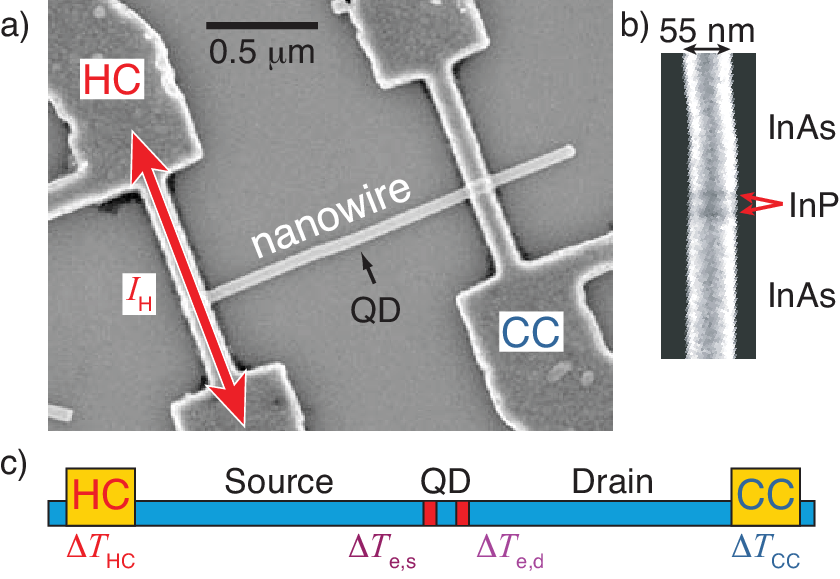}
\caption{(a) A scanning electron micrograph (SEM) image of the HNW system showing the hot (HC) and cold (CC) Au contacts, and the approximate position of the quantum dot (QD), not visible here. A temperature gradient is applied using an AC heating current, $I_\text{H}$, through the HC. (b) Close-up SEM image of the two InP barriers within the InAs nanowire. (c) Cross section of the nanowire and contacts along the nanowire axis. $\Delta T_{\text{(HC,CC)}}$ are the heating-induced electron temperature rises above the background temperature, $T_\text{0}$, in the hot and cold contacts respectively. $\Delta T_{\text{e,(s,d)}}$ are the source and drain electron temperature rises measured in the vicinity of the quantum dot. The source and drain portions of the NW are defined by the direction of heat flow, the source being closest to the HC.}
\label{fig:HNWSchematic}
\end{center}
\end{figure}

To probe e-ph coupling in NWs, we have carried out temperature measurements of the electron temperature profile around a double-barrier heterostructure embedded into an InAs nanowire (Fig.~\ref{fig:HNWSchematic}). When one end of the nanowire is heated, we observe an increase in the electron temperature on the cold side of the nanowire, which is not expected when considering only electronic diffusive heat flow. By combining these measurements with a FEM model of our heterostructure nanowire (HNW) system, we extract information about e-ph coupling, as well as other thermal transport properties, within the InAs portion of the HNW. Specifically, we find that e-ph interaction within the HNW plays a key role in explaining the observed electron temperatures, and that we must consider both the electronic and phononic thermal conductivities.

In the following, we first describe our heterostructure nanowire system and temperature measurements. We follow this with a discussion on our heat flow model for the HNW and its implementation using FEM modeling. Following a discussion of the FEM results, we lay out theoretical calculations of e-ph interaction in InAs NWs. Lastly, we compare our FEM and theory results on e-ph interaction in NWs with literature values.

\section{Experimental Device and Temperature Measurements}
The HNW contains a quantum dot (QD) defined by two 4~nm InP barriers, spaced approximately 14~nm apart (Fig.~\ref{fig:HNWSchematic}(b)). The HNW has an average diameter of 55~nm, a length of 1.26~$\mu$m between its two Ni/Au (25/75 nm) contacts, and an estimated carrier density of $n_{carrier} = 2.6\pm0.3\times10^{17}$~cm$^{-3}$ at 4.2 K. The leads and HNW lie on top of a Si substrate with a 100~nm thick SiO$_2$ capping layer. The measurement set-up and thermometry techniques are described in detail in Refs.~\onlinecite{Hoffmann2009} and \onlinecite{Hoffmann2009_2}. In brief, two 180$^\circ$ out-of-phase AC heating voltages applied to the hot contact (HC) are used to heat the electrons on the source side of the HNW (Fig.~\ref{fig:HNWSchematic}). Note that the heating current ($I_\text{H}$) does not actually flow through the HNW itself. This is achieved by tuning the amplitudes of the two voltages to create an AC voltage node at the location of the HNW, thus keeping the bias voltage due to the heating current at negligible levels. The temperature of electrons entering the QD is determined by comparing temperature-bias-driven current to voltage-bias-driven current.\cite{Hoffmann2009, Hoffmann2009_2} An external bias voltage applied across the HNW controls the direction of current flow, which determines whether $\Delta{}T_\text{e,s}$ or $\Delta{}T_\text{e,d}$ (see Fig.~\ref{fig:HNWSchematic}) is measured. If electrons flow from the HC to the CC, $\Delta T_{\text{e,s}}$ is measured, while electron flow from the CC to the HC measures $\Delta T_{\text{e,d}}$.\cite{Hoffmann2009, Hoffmann2009_2}

\begin{figure}
\begin{center}
\includegraphics{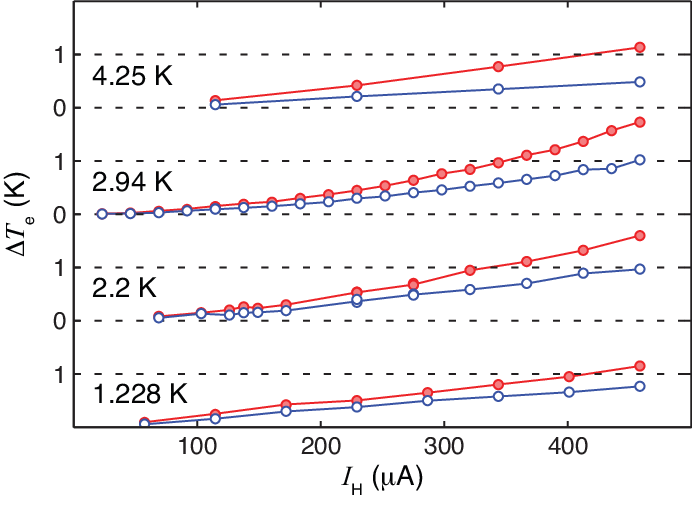}
\caption{Measured source (red filled circles), $\Delta T_{\text{e,s}}$, and drain (blue open circles), $\Delta T_{\text{e,d}}$, electron temperature rises for $T_\text{0}$ = 1.228, 2.2, 2.94, and 4.25 K. The relatively large temperature rise observed for the drain electrons, combined with the low electron conductance of the QD, indicates the presence of phonon-mediated heat flow into the drain electron reservoir.}
\label{fig:ExpdTeVsIH}
\end{center}
\end{figure}

In Fig.~\ref{fig:ExpdTeVsIH} we show measurements of $\Delta T_{\text{e,(s,d)}}(I_\text{H})$ at four different background temperatures, $T_\text{0}$: 1.228, 2.2, 2.94, and 4.25 K. Our HNW has an electrical resistance of 3.2 M$\Omega$, roughly three orders of magnitude higher than in similar InAs NWs without embedded heterostructures. We can therefore assume that the electric resistance is dominated by the QD. Within an electronic diffusive heat model, the QD's high electric resistance would lead one to expect virtually all of the temperature differential applied to the HNW to fall across the QD. If this model were complete, the drain electron temperature should remain at the background temperature, $\Delta T_\text{e,d} \approx 0$. In contrast, for each $T_\text{0}$ we find an unexpectedly warm drain electron temperature, suggesting the presence of at least one significant, additional heat flow mechanism warming the drain electron reservoir. Through FEM modeling, we will show that e-ph coupling within the HNW can explain our key observation of an increased $\Delta T_\text{e,d}$.

\section{HNW Heat Flows and Finite Element Modeling} \label{sec:FEM}
\begin{figure}
\begin{center}
\includegraphics{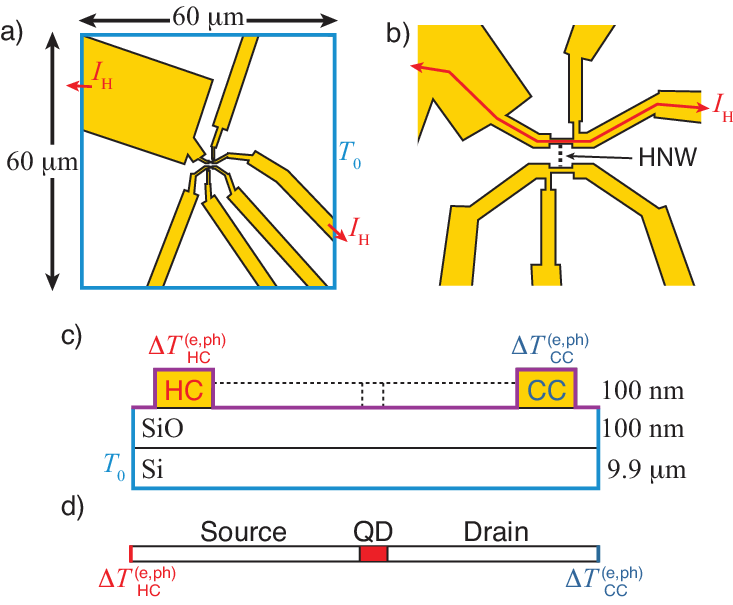}
\caption{(a,b) Top and (c) side view schematics of the simulated area for the Au leads and substrate, called the substrate simulations in the text. The location of where the HNW would be is shown by dashed lines in panels (b) and (c). The outer boundary in (a) and lower boundaries in (c) (blue lines) are set to the background temperature, $T_\text{0}$. The top boundary in (c) (purple line) is assumed to be a perfect insulator. The red line in panel (b) shows the path of the heating current, $I_\text{H}$. Panel (d) shows a schematic of the simulated HNW. In panels (c) and (d), $\Delta T_\text{HC}^\text{(e,ph)}$ and $\Delta T_\text{CC}^\text{(e,ph)}$ represent the electron and phonon temperature rises in the HC and CC, respectively, as well as the boundary conditions for the simulated HNW.}
\label{fig:FEMSchematics}
\end{center}
\end{figure}

To study heat flows through the HNW system, we separately model the HNW and the surrounding leads and substrate using finite element modeling in COMSOL Multiphysics, see Fig.~\ref{fig:FEMSchematics}. We model the electron, $T_\text{e}$, and phonon, $T_\text{ph}$, temperatures using two coupled diffusive heat equations,
\begin{align}
-\vec{\nabla} \cdot \left( \kappa_\text{e}(T_\text{e}) \vec{\nabla}T_\text{e} \right) &= Q_\text{J} - Q_\text{e-ph}\,, \label{eqn:ElectronHeatDiffusion}\\
-\vec{\nabla} \cdot \left( \kappa_\text{ph}(T_\text{ph}) \vec{\nabla}T_\text{ph} \right) &= Q_\text{e-ph}\,. \label{eqn:PhononHeatDiffusion}
\end{align}
Here $\kappa_\text{(e,ph)}$ are the electron and phonon thermal conductivities, respectively, and $Q_\text{J}$ represents Joule heat generated in the electron system. The $Q_\text{e-ph}$ term describes the heat exchanged between electrons and phonons (e-ph interaction) in the Au leads and HNW, and we assume it follows a general power law of the form,
\begin{equation} \label{eqn:Q_eph}
Q_\text{e-ph} = \Gamma \left( T_\text{e}^n - T_\text{ph}^n \right) \,,
\end{equation}
where the parameters $\Gamma$ and $n$ describe the strength and type of e-ph coupling respectively.\cite{Kubakaddi2007}

We simulate the electric potential, $\phi$, through the metallic leads using Laplace's equation,
\begin{equation} \label{eqn:LaplacianVoltage}
\nabla^2\phi=0\,,
\end{equation}
with the primary purpose of determining the amount of Joule heating in the hot contact. However, to determine the amount of Joule heating in the HNW, we use the experimentally measured electric current and electrical conductance of the HNW.

We account for the temperature dependencies of $\kappa_\text{e}$ and $\kappa_\text{ph}$ by using the Wiedemann-Franz law for electrons,
\begin{equation} \label{eqn:ElectricalThermalConductivity}
\kappa_\text{e}(T_\text{e}) = L_\text{0}\sigma T_\text{e}\,,
\end{equation}
and a cubic power law for phonons, valid well below the Debye temperature,
\begin{equation} \label{eqn:PhononThermalConductivityNW}
\kappa_\text{ph}(T_\text{ph}) = C_\text{ph} T_\text{ph}^{3}\,.
\end{equation}
Here, $L_\text{0} = (\pi^2/3)(k_\text{B}/e)^2=2.44\times10^{-8}$~W$\Omega$/K$^2$ is the Lorenz number, $\sigma$ is the electrical conductivity, and $C_\text{ph}$ is a material constant. We assume that the Debye temperatures can be described by their bulk values: $\Theta_\text{D}^\text{Au}\approx163$~K, and $\Theta_\text{D}^\text{InAs}\approx255$~K. In general, $\sigma$ could be temperature dependent,
\begin{equation} \label{eqn:ElectricalConductivity}
\sigma = C_\text{e} f(T_\text{e})\,,
\end{equation}
where $f(T)$ is some temperature dependent, dimensionless function; however, based on electrical conductivity measurements on similar InAs nanowires and thin film Au structures, we assume $f(T)=1$ for the range of temperatures looked at here.

\begin{table}
\begin{center}
\begin{tabular}{|c|c|}
\hline
Parameter & Au \\ \hline
\expandTable{} $C_\text{ph}$ (W/K$^{4}$m) & $3 \times 10^{-3}$$^($\footnote{\label{fn:Cp}Bulk values\cite{Simmons1971} for Au assuming a phonon mean free path of 100~nm, the film thickness.}$^)$ \\ \hline
\expandTable{} $C_\text{e}$ (1/$\Omega$m) & $4\times10^7$$^($\footnote{\label{fn:Ce}Measured separately on a Au thin film structure similar to those used here.}$^)$ \\ \hline
\expandTable{} $\Gamma$ (W/m$^{3}$K$^{5}$)& $1.8\times10^9$$^($\footnote{\label{fn:Gamma}Average of the values from Refs.~\onlinecite{Henny1997, Echternach1992}.}$^)$ \\ \hline
\expandTable{} $n$ & 5$^{(\text{\ref{fn:Gamma}})}$ \\ \hline
\end{tabular}
\end{center}
\caption{Simulation parameters used for the Au leads.}
\label{tab:AuSiSiO}
\end{table}
\begin{table}
\begin{center}
\begin{tabular}{|c|c|c|c|}
\hline
\multirow{2}{*} {Parameter} & Lower & Upper & \multirow{2}{*} {Comments}\\
& Bound & Bound &\\ \hline
$C_\text{ph}$& \multirow{2}{*} {$5\times10^{-5}$} & \multirow{2}{*} {$1.5\times10^{-3}$} & \multirow{2}{*} {See footnote (\ref{fn:CphNW})} \\
(W/K$^{4}$m) &&& \\ \hline
$C_\text{e}$ & \multirow{2}{*} {$5\times10^2$}& \multirow{2}{*} {$5\times10^5$} & Measurements\\(1/$\Omega$m)&&&on pure InAs NWs$^{(\text{\ref{fn:ExtendedRanges}})}$ \\ \hline
$\Gamma$ & \multirow{2}{*} {$10^{6}$} & \multirow{2}{*} {$10^{13}$} & \\
(W/m$^{3}$K$^{n}$)&&& Refs.~\onlinecite{Sugaya2002}, \onlinecite{Prasad2004}, \onlinecite{Prasad2004_2},\\ \cline{1-3}
\expandTable{} $n$ & $0.1$ & $7$ & \onlinecite{Kivinen2004, Henny1997, Steinbach1996, Roukes1985}$^{(\text{\ref{fn:ExtendedRanges},\ref{fn:GammaAndn}})}$\\ \hline
\end{tabular}
\end{center}
\footnotetext{\label{fn:CphNW}Values initially based on Drude model, but were adjusted to conform to a region where the FEM simulations converge.}
\footnotetext{\label{fn:ExtendedRanges}Ranges extended beyond values found in references to encompass possible deviations.}
\footnotetext{\label{fn:GammaAndn}References include examples of bulk, thin film, and NW values.}
\caption{HNW parameter ranges used in the Nelder-Mead optimization method.}
\label{tab:ParameterBounds}
\end{table}

For the surrounding leads to the HNW, we assume that the transport properties are well described by bulk Au, and will stop mentioning the Ni component. The values we use for the four available parameters, \{$C_\text{e}$, $C_\text{ph}$, $\Gamma$, $n$\}, for the Au leads are shown in Table~\ref{tab:AuSiSiO}. Unfortunately, the setup of our system prevents us from measuring any of these four parameters for the InAs portion of the HNW directly. However, based on prior research on similar InAs NWs and e-ph coupling, we can define a range of reasonable values for our HNW, see Table~\ref{tab:ParameterBounds}. Tabulated values from Refs.~\onlinecite{Zeller1971, Ho1972} are used for the Si and SiO$_2$ phonon thermal conductivities. We assume no electric current flows through the Si and SiO$_2$.

During thermometry measurements, a DC bias voltage, $V_\text{B}$, generates current, $I_\text{NW}$, via resonant tunneling through the QD. (Only a single energy level is relevant because the energetic difference between neighboring QD energy levels in this device is sufficiently greater than $k_\text{B}T$ and $eV_\text{B}$ \cite{Hoffmann2009}.) We assume that electrons traversing the QD do so ballistically and elastically. As such, electrons exit the QD into the down-current electron reservoir at an energy higher than the down-current electrochemical potential.\cite{Hoffmann2009_2} Exiting electrons then thermalize within an inelastic mean free path, resulting in Joule heat.\cite{Taboryski1995} This effect is added to our FEM model by uniformly depositing heat, $Q^\text{QD}_\text{J} = I_\text{NW}^2R_\text{QD}$, into the electron system within one electron-electron mean free path\cite{Hansen2005} of the QD. Here, $R_\text{QD}$ is the calculated electrical resistance of the QD, which depends on the measured resistance of the HNW, and $C_e$.

We model the phonons in the QD by calculating an effective phononic thermal conductivity based on the Debye model,\cite{Ashcroft1976} and the geometry of the InP/InAs/InP heterostructure. The acoustic-mismatch model\cite{Swartz1989} predicts a phonon transmission coefficient of 0.999 at each of the InP/InAs interfaces, such that phonon thermal boundary resistances are negligible at these interfaces.

The optimization procedure we lay out in Sec.~\ref{sec:OptimizedParameterSets} requires numerous simulation runs with numerous sets of HNW parameters. Therefore, instead of modeling the Si-SiO$_2$ substrate, Au leads, and HNW all at once, we reduce the computation time by splitting the FEM simulations into two separate sub-models: one for the Si-SiO$_2$ substrate and Au leads, and one for the HNW. The substrate simulations were ran once and used as the boundary conditions for $T_\text{e}$ and $T_\text{ph}$ in the HNW simulations. As a check, a representative set of HNW parameters were simulated both in the HNW model and a comprehensive model that includes the substrate, leads, and HNW. A comparison of these two models shows a difference of roughly 1\%, which validates the use of the two sub-models for the system. Schematics of the two systems with their respective boundary conditions are shown in Fig.~\ref{fig:FEMSchematics}.

We estimate that radiative losses, and direct heat exchange between the leads/HNW and helium (He-3) bath can be neglected when compared to diffusive heat flows and the heat exchanged between the electron and phonon systems. We also note that ballistic electron and phonon effects in the HNW would invalidate Eqs.~\eqref{eqn:ElectronHeatDiffusion} and \eqref{eqn:PhononHeatDiffusion}; however, these effects are beyond the scope of our paper.

\section{Heating the Drain Electrons}
Our model contains three heat flow mechanisms through the HNW system capable of delivering heat to the drain electron reservoir: (i) Joule heat and electron diffusion through the HNW, (ii) phonon diffusion between the HC and CC through the substrate, and (iii) phonon-mediated heat flow through the HNW. As we will show in the following, only mechanism (iii) can account for the observed electron temperatures. 

Our  simulations predict that both mechanisms (i) and (ii) result in $\Delta T_\text{e,d}\approx1$ mK, 2-3 orders of magnitude smaller than seen in our experiments. The magnitude of mechanism (i) is small due to the high electronic thermal resistance of the quantum dot, which prevents heat from reaching the drain electron reservoir. We should note that we expect the Wiedemann-Franz law (WFL) to break down in our QD\cite{Sivan1986, Zianni2007, Hoffmann2009}; however, the effect of mechanism (i) is so small that we can safely ignore Lorenz numbers less than $100L_\text{0}$. The small magnitude of heating due to mechanism (ii) is a result of the low heating power used. Systems where mechanism (ii) is utilized to generate thermal gradients typically have thicker silicon dioxide layers and generate significantly more (at least two orders of magnitude) Joule heat per unit volume in the heating contact.\cite{Small2003}

Our simulations predict that mechanism (iii) is capable of heating the drain electron reservoir by 100s of mK. This mechanism works in three steps to bypass the electrically insulating QD: (1) transfer heat from the hot source electron reservoir to the source phonons though e-ph interaction, (2) heat the drain reservoir phonons through phononic heat diffusion, and finally (3) use the warm drain phonons to heat the drain electron reservoir through e-ph interaction. A similar type of energy transfer process has been previously seen in 2DEG devices.\cite{Schinner2009}

\begin{figure}
\begin{center}
\includegraphics{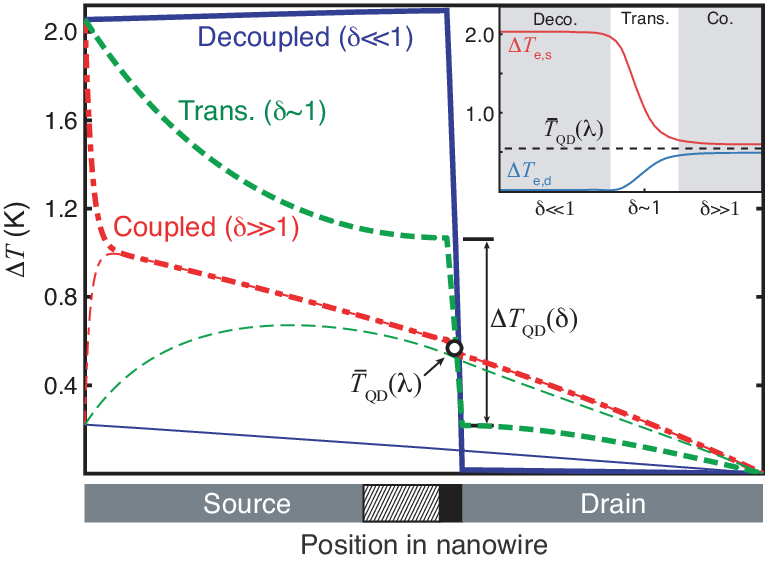}
\caption{Example coupled (dot-dash lines) and decoupled (solid lines) electron (thick) and phonon (thin) temperature profiles through the HNW for a measurement of $\Delta{}T_\text{e,d}$. Also shown is a temperature profile (dashed line) between the coupled and decoupled regimes, the transition regime. For the plots shown here, $\lambda\approx0.9$; note that this value was chosen for demonstration purposes. The hashed region along the HNW position represents the down-current Joule heating due to the voltage drop across the QD. The inset shows how $\Delta{}T_\text{e,(s,d)}$ depend on the degree of coupling.}
\label{fig:dTeProfileVSGamma}
\end{center}
\end{figure}

The effects of each heat flow are illustrated by modeled electron and phonon temperature profiles along the HNW axis; Fig.~\ref{fig:dTeProfileVSGamma} shows three simulated temperature profiles for a drain electron temperature measurement. Here, Joule heating occurs to the left of the QD because for a drain-electron temperature measurement, electrons flow from the drain reservoir to the source reservoir.\cite{Hoffmann2009, Hoffmann2009_2} The three temperature profiles shown are the decoupled and coupled cases, and a transition state. The decoupled state refers to the state where the electron and phonon temperatures are completely independent of each other. Likewise, the coupled state refers to when the electron and phonon systems act as one effective system, $T_\text{e}=T_\text{ph}$. Mechanism (i) results in slight increases in $\Delta{}T_\text{e,(s,d)}$, and is easily seen in the decoupled temperature profile around the QD. Mechanism (ii), seen as an increased $\Delta{}T_\text{CC}$, is too small to be visible in the graph and has a negligible effect on $\Delta{}T_\text{e,d}$. Mechanism (iii) leads to a transfer of heat between the source and drain electron reservoirs, the results of which are the transition and coupled temperature profiles.

\subsection{Two Degrees of Freedom}
To gain a deeper understanding of the factors that determine the temperature profiles, we now look at results from the linearized form of the diffusive heat model defined by Eqs.~\eqref{eqn:ElectronHeatDiffusion} and \eqref{eqn:PhononHeatDiffusion}. Importantly, in this model there are only two physical degrees of freedom that describe how heat flows through the material:
\begin{equation} \label{eqn:delta}
\delta = \sqrt{ \frac{\kappa_\text{e-ph}}{\kappa_\text{e}} + \frac{\kappa_\text{e-ph}}{\kappa_\text{ph}} }
\end{equation}
where
\begin{equation}
\kappa_\text{e-ph} \equiv \Gamma n T_\text{0}^{n-1}d^2\,,
\end{equation}
and
\begin{equation} \label{eqn:ThermalConductivityRatio}
\lambda \equiv \frac{\kappa_\text{e}}{\kappa_\text{ph}}\,.
\end{equation}
Here $d$ is a characteristic length scale, see next section, and $\kappa_\text{e-ph}$ can be interpreted as an effective thermal conductivity between the electron and phonon systems due to e-ph interaction. Note that both $\delta$ and $\lambda$ are dimensionless quantities, and that $\delta$ is derived by solving the linearized forms of Eqs.~\eqref{eqn:ElectronHeatDiffusion} and \eqref{eqn:PhononHeatDiffusion}. $\delta$ represents the degree of thermal coupling between the electron and phonon systems: $\delta\ll1$ represents the decoupled state, while $\delta\gg1$ the coupled state. See the Appendix for further details on how $\delta$ and $\lambda$ are used to match the simulated $\Delta T_\text{e,(s,d)}$ to the observed values.

Since we are interested in matching our simulations to two experimental temperatures, $T_\text{e,(s,d)}$, $\delta$ and $\lambda$ are the two parameters that we ultimately want to manipulate. Due to this limited number of relevant parameters, multiple combinations of the original four parameters, \{$C_\text{e}$, $C_\text{ph}$, $\Gamma$, $n$\}, exist that will result in the same $\delta$ and $\lambda$ pair. As such, we cannot expect to extract unique values for all four physical parameters.

\section{Optimized parameter sets} \label{sec:OptimizedParameterSets}
We search for optimized parameter sets that minimize the error between the simulated electron temperatures and the experimental data by repeatedly applying the Nelder-Mead optimization method\cite{Nelder1965}. Each time the method is run, the starting Nelder-Mead simplex is initialized with random values chosen from within the parameter bounds defined in Table~\ref{tab:ParameterBounds}, but is then free to explore an unbounded parameter space for the duration of the optimization run. We define the simulation error as the RMS difference between the simulated and measured temperatures,
\begin{equation}
\epsilon_\text{error} \equiv \sqrt{ \frac{1}{N} \sum_i \left( T_\text{e, sim.}^{(i)} - T_\text{e, exp.}^{(i)}\right )^2} \,,
\end{equation}
where $i$ denotes the $i$'th experimental heating current, and $N$ is the number of heating currents used in the optimization procedure for each $T_\text{0}$. We use $N=4$  and fit to the $I_\text{H}<250$ $\mu$A portions of the experimental curves in Fig.~\ref{fig:ExpdTeVsIH}, except for $T_\text{0}=4.25$ K where we use all four data points. After finding numerous parameter sets that fit our data for each $T_\text{0}$, we find two relevant trends between the four parameters: one between the conductivity coefficients, $\{C_\text{e},C_\text{ph}\}$, and one between the two e-ph coupling parameters, $\{n,\Gamma\}$. We also find interesting trends in the calculated $\delta$ and $\lambda$ values. These trends are discussed in the following.

\subsection{Optimized Conductivity Parameters}
\begin{figure}[t]
\begin{center}
\includegraphics{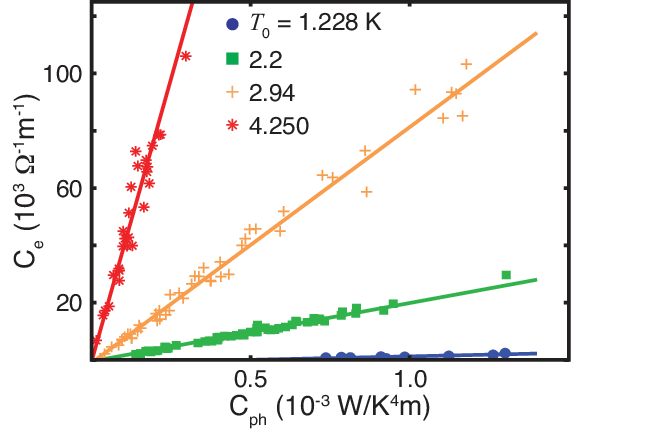}
\caption{Scatter plot of the optimized $C_\text{e}$ and $C_\text{ph}$ parameters with linear fits to each $T_\text{0}$ parameter set.}
\label{fig:Optimized}
\end{center}
\end{figure}

Here we consider two calculated values from the optimized parameters sets: the ratio of the two thermal conductivity parameters, $C_\text{e}$ and $C_\text{ph}$, and the calculated $\lambda$ values, the former is plotted in Fig.~\ref{fig:Optimized}. In both cases we find a temperature dependence that scales as a power law in $T_\text{0}$:
\begin{equation} \label{eqn:CeCpRatio}
\frac{C_\text{e}}{C_\text{ph}} = (9.55\times10^{6}\text{ W}^{-1}\Omega^{-1}\text{K}^{-0.1})T_\text{0}^{4.1}\,,
\end{equation}
and
\begin{equation} \label{eqn:LambdaResult}
\lambda = (0.113\text{ K}^{-2.7})T_\text{0}^{2.7}\,.
\end{equation}

Equation~\eqref{eqn:CeCpRatio} demonstrates the presence of an unaccounted-for temperature dependence in our model, since $C_\text{e}$ and $C_\text{ph}$ were originally assumed to be temperature independent. Without further measurements however, it is unclear whether this temperature dependence comes from the electron system, the phonon system, or a combination of the two.

Equation~\eqref{eqn:LambdaResult} implies that the electron and phonon contributions to the total thermal conductivity in the HNW are comparable between 1.2 and 4.2~K. Additionally, the electronic contribution is the larger of the two for most of this temperature range, with a crossover temperature of roughly 2.2~K.

One possibility that may help explain these results is a breakdown of the WFL in the InAs portion of the HNW. We can check the WFL for our HNW by inserting Eqs.~\eqref{eqn:CeCpRatio} and \eqref{eqn:LambdaResult} into Eq.~\eqref{eqn:ThermalConductivityRatio} and solving for the Lorenz number, L:
\begin{align}
\lambda &= \frac{LC_\text{e}}{C_\text{ph}} T_\text{0}^{-2}\nonumber \\
L &= L_\text{0}(0.486\text{ K}^{-0.6})T_\text{0}^{0.6}\,.
\label{eqn:LorBD}
\end{align}
The apparent temperature dependence of $L$ suggests that the WFL does indeed break down in the InAs portion of the HNW. Since heat transfer between electrons and phonons is the key component to transferring heat between the two electron reservoirs in our HNW, it may follow that inelastic e-ph scattering is the cause of the breakdown; however, the breakdown could also be due to other effects, such as changes in the Fermi energy.

The last point we note about the thermal conductivity parameters is that the slopes of the four data sets in Fig.~\ref{fig:Optimized} are dependent on $f(T_\text{e})$, $L$, and $n_\text{ph}$. If we change any of these parameters to remove the observed temperature dependence in the slopes, only the coefficients in Eqs.~\eqref{eqn:CeCpRatio}-\eqref{eqn:LorBD} change; the general results of our paper remain the same.

\subsection{Optimized Electron-Phonon Parameters}
\begin{figure}
\begin{center}
\includegraphics{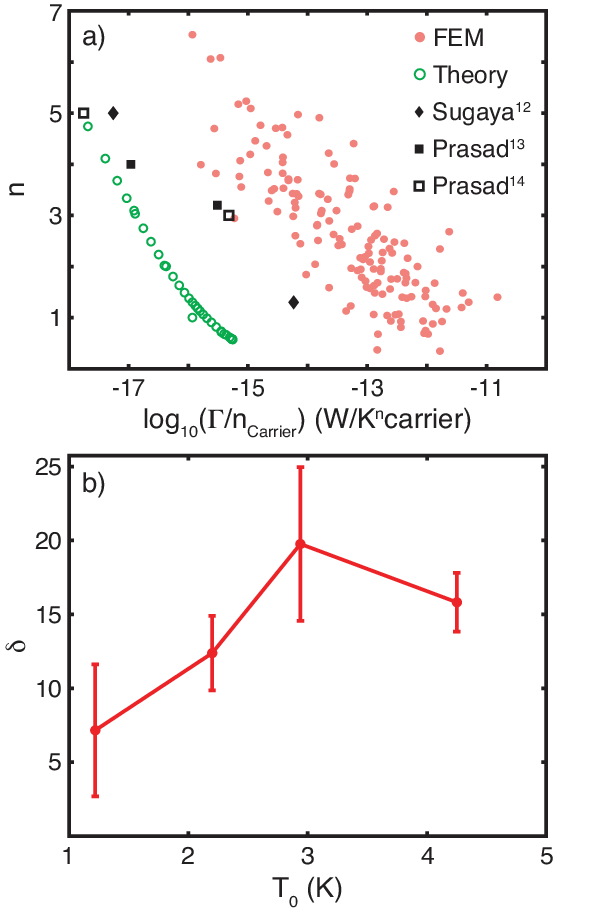}
\caption{(a) Comparison of the optimized $n$ and $\Gamma$ parameter sets from the FEM simulations, theoretical calculations, and literature values\cite{Sugaya2002, Prasad2004, Prasad2004_2}. The $\Gamma$ values here have been normalized by the carrier density to allow for a comparison similar to the power loss per carrier. The theoretical data shown here include both deformation and piezoelectric e-ph scattering; each theory data point represents a different electron temperature between 0.7 and 4.7~K and a background temperature of 2.94~K. (b) Scatter plot of the $\delta$ values, calculated from the optimized FEM parameter sets, as a function of $T_\text{0}$.}
\label{fig:nVSGammaNormalized}
\end{center}
\end{figure}

The relation between $n$ and $\Gamma$ resulting from our model is shown in Fig.~\ref{fig:nVSGammaNormalized}(a). Our results show a band of values, from which no particular pair of $n$ and $\Gamma$ can be specified. Additionally, we cannot use the simulation error to pick out any regions along the band since no minimum exists.

Though we cannot extract unique values for $n$ and $\Gamma$, we can estimate the degree of coupling, $\delta$ by inserting each optimized parameter set into Eq.~\eqref{eqn:delta}. Since we are interested in e-ph coupling over the length scale of the InAs portions of the HNW, we chose $d$ to be half the length of the HNW, $d=630$~nm. Figure~\ref{fig:nVSGammaNormalized}(b) shows the calculated $\delta$ values as a function of $T_\text{0}$. The range of $\delta$ values implies that the electrons and phonons in the HNW are essentially in the fully coupled regime.

\section{Theory}
We compare the optimized parameter sets extracted from our simulations with theoretical equilibrium power loss calculations in a nanowire. As a model, we use a cylindrical nanowire of radius $R = 25$~nm with an infinite potential well in the radial direction, assuming plane waves with a parabolic dispersion along the nanowire axis:
\begin{align}
\varphi_{\nu j k}(r,\varphi) &= \frac{1}{\sqrt L} e^{i k z} R_{\nu j}(r) e^{i \nu \varphi}\,, \\ 
R_{\nu j}(r) &= \frac{J_\nu(\alpha_{\nu j} \frac{r}{R})}{\sqrt{\pi} R J_{\nu+1}(\alpha_{\nu j})}\,,
\end{align}
where $\alpha_{\nu j}$ is the j-th zero of the Bessel functions $J_\nu(x)$. In order to determine the chemical potential and thus the occupied electronic sub-bands, the three-dimensional doping density is translated into the one-dimensional electronic density: $n_{1d} = n_{3d} \pi R^2$. With this, we can calculate the average power loss rate per electron\cite{DasSarma1993}
\begin{equation} \label{eqn:TheoryPower}
\tilde{P}(T_e) = \left\langle \frac{dE}{dt} \right\rangle = \frac{1}{N} \sum\limits_{\bf q} \hbar \omega_{\bf q} \frac{dN_{\bf q}}{dt}\,,
\end{equation}
where $\hbar \omega_{\bf q}$ is the phonon energy, $N_{\bf q}$  is the occupation of the phonon mode ${\bf q}$, and $N$ is the number of electrons in the normalization volume considered. The transition rates $dN_{\bf q}/dt$ are evaluated by Fermi's golden rule for the scattering mechanisms addressed below.

For the interaction of the electronic system with the lattice, that is, the e-ph interaction, 
\begin{equation}
H = \sum\limits_{\nu' j' k'} \sum\limits_{\nu j k} a_{\nu ' j' k'}^{\dagger} a_{\nu j k} \sum\limits_{\bf q} g_{\nu ' j', \nu j}^{\bf q} (b_{\bf q}^{\dagger} + b_{-\bf{q}}^{\,})\,,
\end{equation}
we consider confined one-dimensional phonons corresponding to the modes present in an isolated nanowire. These are calculated within an isotropic continuum model following Ref.~\onlinecite{Stroscio1994, Stroscio2001}, and assuming free-surface boundary conditions. This results in a multitude of quantized phonon modes $ \omega_{\bf q \kappa}$ with $\kappa$ labeling the individual modes.

We consider only compressional modes in the calculations; for the deformation potential coupling, torsional modes do not couple. Flexural modes in principle couple to higher, azimuthally dependent states, and are expected to increase the power loss beyond the values calculated here. For piezoelectric coupling, the piezoelectric constants for wurtzite InAs are not known. We thus calculate the constants from the zinc-blende value, $e_{14}$, using the transformation described in Ref.~\onlinecite{Larsson2007}. There, a zinc-blende lattice in the (111) growth direction is considered, which corresponds to the wurtzite lattice in a nearest neighbor approximation.  The parameters used for our calculations are the lattice temperature $T_\text{0} = 2.94$~K, the deformation potential of the conduction band $D = -5.08$ eV, the mass density $\rho = 5680$~kg/m$^3$, and the longitudinal speed of sound $v_l = 4410$~m/s. For more details, see Ref.~\onlinecite{Weber2010}. We note that this theory predicts equal contributions from the deformation and piezoelectric e-ph couplings. In bulk materials, optical phonons are typically assumed to contribute negligible amounts due to being frozen out.

Using Eq.~\eqref{eqn:TheoryPower} in combination with Eq.~\eqref{eqn:Q_eph}, we can convert $\tilde{P}(T_\text{e})$ to an $n$ vs $\Gamma$ plot,
\begin{align}
n &= T_\text{e} \frac{ \partial^2 \tilde{P}/\partial T_\text{e}^2}{ \partial \tilde{P}/\partial T_\text{e}} + 1\,,\label{eqn:nFit} \\
\Gamma &= \frac{\tilde{P}(T_\text{e})}{T_\text{e}^n-T_\text{ph}^n}\,. \label{eqn:GammaFit}
\end{align}
The result of these calculations are shown in Fig.~\ref{fig:nVSGammaNormalized}(a) alongside our FEM results.

Interestingly, the shape of the curve mapped out by the theory values closely follows the shape of our FEM results instead of localizing about a single $(n,\Gamma)$ pair. The range of $n$ and $\Gamma$ values in both data sets, particularly the theory data set, suggests that e-ph interaction in our HNW does not follow the power law equation given by Eq.~\eqref{eqn:Q_eph}; previous research on similar NWs found an exponential dependence\cite{Prasad2004}. Also of note is that the theory values are around $10^3$ times smaller than the FEM results. This observation, that deformation potential scattering with acoustic phonons underestimates the energy transfer, goes well with previous transport studies in gated nanowires\cite{Weber2010}.

\section{Comparison with literature values}
Finally, we compare our results to the electron energy power loss experiments conducted by Sugaya et al.\cite{Sugaya2002} and Prasad et al.\cite{Prasad2004,Prasad2004_2} Note that the wires in both of these papers were constructed out of InGaAs 2DEG materials, compared to our vertically grown HNW. The 2DEG wires in these papers range from 25 to 770 nm wide and 8 to 25 nm thick.\cite{Sugaya2002, Sugaya2002_2, Sugaya2002_3, Prasad2004, Prasad2004_2} The $n$ and $\Gamma/n_{carrier}$ values from their data are shown in Fig.~\ref{fig:nVSGammaNormalized}(a).

To within a factor of 10, these values agree well with our parameter sets, with our band of values being the larger. Theoretical calculations by Vartanian et al.\cite{Vartanian2008} have recently shown that increasing the degree of lateral confinement in NWs enhances the electron energy loss rate. Since our HNW has a larger degree of confinement, due to being vertically grown instead of etched out of a 2DEG, it should then exhibit larger electron energy loss rates, which is exactly what we see in Fig.~\ref{fig:nVSGammaNormalized}(a).

It is interesting to note that, like our theory values, the literature values lie along a band similar in structure to our FEM results. Whether this is indicative of an underlying physical trend, or merely a consequence of the equational form of Eq.~\eqref{eqn:Q_eph} is beyond the scope of this work.

\section{Discussion and Conclusion}
We have shown that by combining FEM with experimental temperature measurements, we can investigate the thermal properties of InAs nanowires, including e-ph interaction. We find that the polynomial form for e-ph interaction given by Eq.~\eqref{eqn:Q_eph} doesn't match our theory calculations, earlier experimental results, nor our data. We have also demonstrated that between 1.2 and 4.2 K, the electrons and phonons appear to be in a coupled state with comparable thermal conductivities.

The large difference between the FEM and theory data sets in Fig.~\ref{fig:nVSGammaNormalized}(a) demonstrates that additional thermal effects, either in our FEM model or in our theoretical calculations, are missing. This is true even with the theoretical result of enhanced e-ph scattering rates due to piezoelectric scattering, as observed in  previous experiments\cite{Weber2010}. In our FEM model, we have neglected thermal boundary resistances, electron-phonon Kapitza conductances\cite{Huberman1994, Sergeev1998, Mahan2009}, and non-equilibrated electron and phonon distributions. Due to our system being in a relatively coupled state, these additional thermal effects could affect the extracted $\delta$ and $\lambda$ values, and thus the $(n,\Gamma)$ band in Fig.~\ref{fig:nVSGammaNormalized}(a), reported here. From the theory side, we previously noted that the inclusion of flexural phonon modes in the HNW could increase the predicted coupling strength. It is unlikely that any single effect will reconcile the roughly three orders of magnitude difference between the theory and simulation data in Fig.~\ref{fig:nVSGammaNormalized}(a). Instead, a combination of additional thermal effects in the FEM model and refining the theory calculations is likely required.

\section*{Acknowledgements}
We acknowledge financial support from NSF IGERT grant No.~DGE-0549503, the National Science Foundation Grant No.~DGE-0742540, ARO Grant No. W911NF0720083, Energimyndigheten Grant No.~32920Ð1, nmC@LU, ESF Research Network EPSD, and the Foundation for Strategic Research (SSF). Effort sponsored by the Air Force Office of Scientific Research, Air Force Material Command, USAS, under grant number FA8655-11-1-3037. The U.S. Government is authorized to reproduce and distribute reprints for Governmental purposes notwithstanding any copyright notation thereon.

\section*{Appendix}
Here we look in greater detail at how $\delta$ and $\lambda$ affect the electron and phonon systems, as well as how they can be used to match the simulated $\Delta T_\text{e,(s,d)}$ values with experiments.

If we interpret the $\kappa_i$'s in Eq.~\eqref{eqn:delta} as heat fluxes instead of conductivities, $\delta$ can be thought of as a comparison of the rate of heat exchange through e-ph interaction, to the rates at which the electron and phonon systems individually dissipate heat to their surroundings through diffusion. If either system dissipates heat slower than the rate of e-ph heat exchange, $\kappa_\text{e-ph}/\kappa_\text{(e,ph)}\gg1$, then that system can be considered coupled. For example, if one has $\kappa_\text{e-ph}/\kappa_\text{e}\gg1$ and $\kappa_\text{e-ph}/\kappa_\text{ph}\ll1$, $T_\text{e}$ will adjust to match $T_\text{ph}$. Meanwhile, $T_\text{ph}$ will remain unchanged from its decoupled state since the phonon system can distribute heat much faster than the e-ph energy exchange rate. As this example demonstrates, one could conceivably have the case where one system's temperature behaves as though it is coupled, while the other behaves as though it is decoupled.

Alternatively, if both the electron and phonon system can be considered coupled, $\kappa_\text{e-ph}/\kappa_\text{e}\gg1$ and $\kappa_\text{e-ph}/\kappa_\text{ph}\gg1$, $\lambda$ can be used to determine which, if either, of the two systems will behave in an uncoupled manner. For example, consider a point along a wire where $\lambda\ll1$, $\kappa_\text{e-ph}/\kappa_\text{e}\gg1$, and $\kappa_\text{e-ph}/\kappa_\text{ph}\gg1$. If $T_\text{e}>T_\text{ph}$, say due to Joule heating, then the net effect of e-ph interaction will be for the electron system to rapidly transfer heat to the phonon system. Since $\kappa_\text{e}\ll\kappa_\text{ph}$, the phonon system will be able to dissipate its gained heat to the surrounding material at a faster rate than the electron system can replenish its lost heat from any heat sources. A similar argument applies to the case for $\lambda\gg1$, where $T_\text{e}$ will remain unchanged from its decoupled state, and $T_\text{ph}$ will adjust to match $T_\text{e}$. For $\lambda$ values near unity, in a homogeneous material the electron and phonon temperatures couple to an effective temperature, $T_\text{eff}$, that lies between the two uncoupled temperatures: $T_\text{(e,ph)}|_{\delta=0} \leq T_\text{eff} \leq T_\text{(ph,e)}|_{\delta=0}$.

For our HNW system, in combination with the four externally controlled thermal boundary conditions: $\{\Delta T_\text{e,HC}, \Delta T_\text{e,CC}, \Delta T_\text{ph,HC}, \Delta T_\text{ph,CC}\}$, $\delta$ and $\lambda$ determine the balance of heat flowing between the source and drain electron reservoirs via the phonon system with the heat gained and lost through electron diffusion to the Au leads. Properly tuned, the two parameters allow the simulated electron temperatures around the QD, $\Delta T_\text{e,s}$ and $\Delta T_\text{e,d}$, to be matched to the experimental measurements. More specifically, $\lambda$ effectively determines the average temperature of the QD in the fully coupled state, $\bar{T}_\text{QD} = (T_\text{e,s} + T_\text{e,d})/2$, while $\delta$ determines the temperature drop across the QD, $\Delta T_\text{QD} = \Delta T_\text{e,s} - \Delta T_\text{e,d}$. Both of these effects can be seen in Fig.~\ref{fig:dTeProfileVSGamma} and its inset.

\end{document}